\documentclass[a4paper,preprintnumbers,showpacs,twocolumn,superscriptaddress,nofootinbib,amsmath,amssymb,notitlepage]{revtex4-1}

\usepackage{enumitem}
\usepackage{times}
\usepackage{dcolumn}
\usepackage{mathrsfs}
\usepackage{comment}
\usepackage{hyperref}
\usepackage{amsmath}
\usepackage{mathtools}
\usepackage{bbm}
\usepackage{bm}
\usepackage{amsfonts}
\usepackage{amssymb}
\usepackage{mathrsfs}
\usepackage{wasysym}
\usepackage{graphicx}
\usepackage[]{subfigure}
\usepackage{filecontents}
\usepackage[dvipsnames]{xcolor}
\usepackage{bbold}

\hypersetup{
    bookmarks=true,         
    pdftoolbar=true,        
    pdfmenubar=true,        
    pdffitwindow=true,     
    pdfstartview={FitH},     
    pdftitle={My title},     
    pdfauthor={author},      
    pdfsubject={Subject},    
    pdfcreator={Creator},    
    pdfproducer={Producer},  
    pdfkeywords={keyword1} 
    pdfnewwindow=true,       
    colorlinks=true ,        
    linkcolor=blue  ,        
    citecolor=blue,      
    filecolor=green,       
    urlcolor=blue            
}

\newcommand{\ed}{\mathrm{d}}

\newcommand{\car}{Carath\'{e}odory}

\newcommand{\Qr}{$\delta Q_\mathrm{rev}$}

\begin{document}

\title{Carathéodory's thermodynamics of the Schwarzschild black hole surrounded by quintessence}

\author{Mohsen Fathi}
\affiliation{Departamento de F\'{i}sica, Universidad de Santiago de Chile, Avenida V\'{i}ctor Jara 3493,    Santiago, Chile}

\author{Mart\'{i}n Molina}
\affiliation{Instituto de F\'{i}sica, Pontificia Universidad Cat\'{o}lica de Valparaíso, Avenida Brasil 2950, Valparaíso, Chile}

\author{J.R. Villanueva}
\email{jose.villanueva@uv.cl}\thanks{Corresponding author}
\affiliation{Instituto de F\'{i}sica y Astronom\'{i}a, Universidad de Valpara\'{i}so,
Avenida Gran Breta\~{n}a 1111, Valpara\'{i}so, Chile}


\begin{abstract}

In this paper, we apply the \car's method of geometrothermodynamics to investigate the behavior of the main thermodynamic parameters associated with a Schwarzschild black hole surrounded by quintessence. The corresponding Pfaffian form is constructed by means of the Schwarzschild radius $r_s$, and the quintessential radius $r_{\gamma}$, as independent variables. This form is then used to characterize the thermodynamic manifold. The homogeneity of the system allows for the recognition of the empirical temperature and entropy, and thus, connects with the usual laws of thermodynamics.
In particular, we show that the Helmholtz and Gibbs free energies lead to the same value for the Schwarzschild black hole, in the case of the vanishing cosmological term.

\bigskip

{\noindent{\textit{keywords}}: Adiabatic processes, black hole thermodynamics, quintessence
}\\

\noindent{PACS numbers}: 04.20.Fy, 04.20.Jb, 04.25.-g   
\end{abstract}

\maketitle

\tableofcontents

\section{Introduction and Motivation}\label{sec:intro}

Attributing the laws of thermodynamics to black holes, as proposed in the 1970's in a series of seminal papers \cite{Bardeen:1973gs,Bekenstein:1972,Bekenstein:1973,Bekenstein:1974,Bekenstein:1975,Hawking:1974sw}, has opened new gates in the study of these mysterious objects, from both theoretical and astrophysical points of view. As an example, and from the theoretical side, the concept of entropy is applied to black holes by means of the famous Bekenstein-Hawking (B-H) entropy formula, although its direct application to the extremal black holes (EBHs) is controversial because the zero entropy conjecture for EBHs \cite{Teitelboim:1994,Carroll:2009}, neglects the direct relationship between the entropy and the event horizon's area. Regarding the experimental and observational evidences, the technological advancements have then provided facilities to do some tests on analogue models, in order to unravel the links between thermodynamic entities and black hole dynamics \cite{Unruh:1981,Novello:2002,Schutzhold:2005,Carusotto:2008,Belgiorno:2010,Weinfurtner:2011,Castelvecchi:2016,Steinhauer:2016,Lima:2019,Kolobov:2021}. 

It is, however, worth mentioning that the axiomatic settings of the laws of classical thermodynamics, themselves, are not always given in the conventional ways. In fact, modern thermodynamics is the result of a long-term process, in which, the laws have been given rigorous tests and undergone different experiments (see the reviews in Refs.~\cite{Hutter:1977,Neumaier:2007}). Among all, the geometric approach proposed by \car~in Ref.~\cite{caratheodory09}, is significant (see the reviews in Refs.~\cite{Landsberg:1956,Antoniou:2002}). Together with the approach given by Gibbs \cite{Gibbs:1949}, the \car's method and its further developments by Born~\cite{Born:1949}, form the foundations of the so-called \textit{Geometrothermodynamics} \cite{Quevedo:2007}. The link between the methods of \car~and Gibbs, is however, argued to be established in terms of the homogeneity of the Pfaffian form \Qr, as the infinitesimal heat exchange reversibly \cite{Belgiorno:2002}, and this method has been applied to the laws of black hole thermodynamics in Refs.~\cite{Belgiorno:2002iv,Belgiorno:2002iw,Belgiorno:2003a,Belgiorno:2003b,Belgiorno:2004}, and recently, in Refs.~\cite{Molina:2021,Fathi:2021EPJC,Fathi:2021PLB} regarding the adiabatic (isoareal) processes of Hayward and BTZ BHs. The geometric formulation of the \car's~method, makes it possible to have a self-contained study of the black hole thermodynamics, by using only the respected spacetime structure. 

If the spacetime is coupled with cosmological parameters, its thermodynamics also reveals the evolutionary structure of the universe, in which the black hole resides. This problem is of our interest in this paper. As it is elaborated in the next section, we take into account a black hole spacetime that is coupled with a quintessential dark field and we apply the \car~thermodynamics to explore the possible adiabatic processes, based on the solutions to the Pfaffian in the context of the black hole geometry. {To elaborate this, we calculate the analytical solutions to the corresponding Cauchy problem, and accordingly, we determine the allowed physical paths on the thermodynamic manifold.}

{The paper is organized as follows: In Sect.~\ref{sec:BlackHole}, we introduce the spacetime and its causal structure. In Sect.~\ref{sec:CaraThermo},  we introduce the \car~geometrothermodynamics and express the black hole thermodynamic parameters as functions of its spacetime components. In this section, we calculate the geometric entropy and temperature, which lead to the interpretation of the Pfaffian in the context of the first law. These parameters are then applied in order to demonstrate the entropy-temperature  behavior within the thermodynamic foliations. In Sect.~\ref{sec:Extremal}, we present the Cauchy problem in the context of isoareal processes, to find the permissible trajectories for adiabatic processes. In Sect.~\ref{sec:HeatCap}, the previously calculated thermodynamic parameters are applied to calculated the heat capacity and the free energies of the black hole, and the appropriate limits are discussed. We conclude in Sect.~\ref{sec:conclusions}.
Throughout this work, we apply a geometrized system of units, in which $G=c=1$.}

\section{The black hole solution in the dark background}\label{sec:BlackHole}

The current study is aimed at the continuation of applying the \car's method to black hole thermodynamics. To elaborate this purpose, we however, choose to include the cosmological dynamics that feature as an evolutionary characteristic of black hole geometries. This way, one needs to take into account the effects from the dark side of the universe. Such features have been discussed extensively in general relativity and alternative gravity theories (see for example Ref.~\cite{JimenezMadrid:2005,Jamil:2009,Li:2019,Roy:2020}). These may include the presence of a dark matter halo \cite{Xu:2018,Das:2021}, or coupling with a quintessential field \cite{Kiselev:2003,Saadati:2019,AliKhan:2020,CFOV21}. 
The standard thermodynamics of static black holes in quintessence has been studied extensively, for example in Refs.~\cite{Tharanath:2013,Ghaderi2016,Ma:2017,Ghosh:2018,Rodrigue:2018,Nam:2020,Lutfuoglu:2021}. We are, however, interested in investigating the geometrothermodynamics of such a black hole (a Schwarzschild black hole) through the \car's method, in order to find the limits imposed on the corresponding thermodynamic manifold, that are not accessible by adiabatic processes.

The static, spherically symmetric black hole solution associated with quintessence 
is described by the line element 
\begin{equation}\label{eq:metric}
    \ed s^2 = -B(r) \ed t^2 + B^{-1}(r) \ed r^2+r^2 \ed\theta^2+r^2\sin^2\theta \ed\phi^2
\end{equation}
in the $x^\mu = (t,r,\theta,\phi)$ coordinates, where the lapse function is given by \cite{Kiselev:2003}
\begin{equation}\label{eq:lapse}
    B(r) = 1
    - \frac{r_s}{r}-\frac{\gamma}{r^{3w_q+1}},
\end{equation}
with $r_s=2M$,
$\gamma$ and $w_q$, representing the parameters of 
quintessence and the equation of state (EoS), and $M$ is the black hole's mass. For an accelerating universe, the EoS parameter respects the range $-1<w_q<-\frac{1}{3}$, and the particular case of $w_q=-1$ recovers the cosmological constant. In this paper, we confine ourselves to  the case of $w_q = -\frac{2}{3}$, that recovers
\begin{equation}\label{eq:lapse_1}
    B(r) = 1
    - \frac{r_s}{r}-\gamma r,
\end{equation}
 and accordingly, $\gamma$ has dimension of $L^{-1}$. 
Now let us recast the line element as
\begin{equation}\label{eq:lapse_2}
    B(r) = 1 - \frac{r_s}{r}- \frac{r}{r_{\gamma}},
\end{equation}
by defining $r_{\gamma}\doteq \frac{1}{\gamma}$ as the quintessential radius. The black hole horizons that are located at the radial distances $r_h$ at which $B(r_h) = 0$, are therefore given by
 \begin{eqnarray}
&& r_{++} =\frac{r_{\gamma}}{2}\left(1 + \sqrt{1-\frac{4 r_s}{r_{\gamma}}}\right),\label{eq:rg_sB}\\
&& r_+ = \frac{r_{\gamma}}{2}\left(1- \sqrt{1-\frac{4 r_s}{r_{\gamma}}}\right),\label{eq:rH_sB}
\end{eqnarray}
which are, respectively, the (quintessential) cosmological and the event horizons. Hence, the extremal black hole, for which $r_+=r_{++}=r_e = 2r_s$, corresponds to the case of $r_\gamma = r_{\gamma e} = 4r_s $, and a naked singularity is occurred when $r_\gamma < r_{\gamma e}$.

\section{The \car~thermodynamics applied to the black hole}\label{sec:CaraThermo}

The \car's~framework of thermodynamics is based on the \car's principle which reads: {\it in the neighbourhood of any arbitrary state $J$ of a thermally isolated system $\Sigma$, there are states $J'$ which are inaccessible from $J$} \cite{chandrasekhar39,Buchdahl:1966,adkins_1983}. This inaccessibility may be established on the integrability of the  appropriate Pfaffian form \Qr \,for the system, which implies that  it can be written as
\begin{equation}
    \delta Q_{{\rm rev}}=\tau \ed\sigma,
    \label{eq:Pffo0}
\end{equation}
where $\tau$ is an integrating factor which is considered to be the empirical temperature, and $\sigma$ is the empirical entropy \cite{chandrasekhar39,Buchdahl:1966,adkins_1983}. The existence of an integrating factor ensures the existence of an infinite number of them. So if the integrating factor is considered to be the absolute temperature $T$, then the Pfaffian form reads
\begin{equation}
    \label{Pffo1} \delta Q_{{\rm rev}}= T {\rm d}S,
\end{equation}
where $S$ is the metric entropy, that is related to the second law for the irreversible processes. The integrating factor can be calculated, once \Qr~represents a symmetry. In this sense, the  thermodynamics manifold is constructed by the foliation of adiabatic hyper-surfaces on which, \Qr$=0$. Therefore, the homogeneity properties of the Pfaffian form allows to connect the Carath\'{e}odory's framework with that of Gibbs \cite{Belgiorno:2002,Belgiorno:2002iv}, and then, all thermodynamics can be applied to obtain the relevant macroscopic magnitudes. Thus, the black hole thermodynamics can be studied using this approach which leads to interesting properties.

In particular, our interest is in the study of the system characterized by the variables $(r_s,r_{\gamma})$, and hence, are chosen as the independent extensive variables of the thermodynamic manifold with the constraint  $1-\frac{4r_s}{r_{\gamma}}>0$, as inferred from Eqs.~\eqref{eq:rg_sB} and \eqref{eq:rH_sB}. Accordingly, this manifold is bounded by the extremal sub-manifold, corresponding to $r_{\gamma}>4r_s$.

{With this in mind, we postulate that the Pfaffian form $\delta Q_{{\rm rev}}$ can be written as
\begin{equation}\label{eq:1stLaw}
    \delta Q_{\mathrm{rev}} = \ed r_s -\Gamma\ed r_{\gamma},
\end{equation}
in terms of the system variables $(r_s, r_\gamma )$, where $\Gamma\equiv\Gamma(r_s,r_\gamma)$ is regarded as the "generalized force" associated with the quintessence contribution. This coefficient is supposed to be non-zero everywhere on the thermodynamic domain, so that the Pfaffian is always non-singular, and its integrability is guaranteed by the condition \Qr$\wedge\ed$(\Qr)$ \,= 0$ \cite{Belgiorno:2002iv}.} In this way, the $\mathrm{d}r_s$ term in Eq. \eqref{eq:1stLaw} plays the role of the infinitesimal changes in the black hole's internal energy, whereas $-\Gamma \mathrm{d}r_\gamma$ is a work term which is identified, completely, in geometrical contexts.

\begin{figure}[h]
    \centering
    \includegraphics[width=9cm]{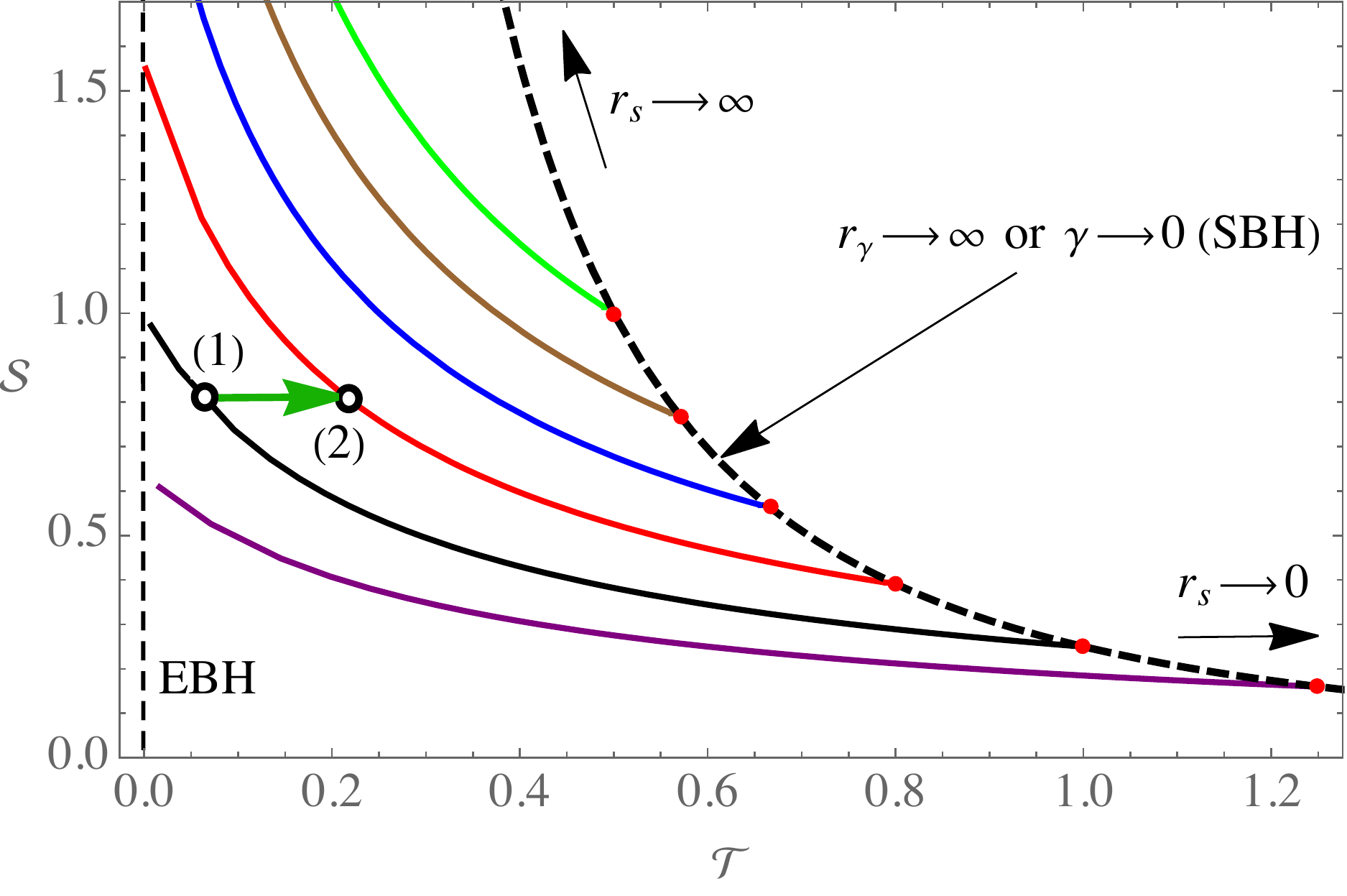}
    \caption{{The $\mathcal{S}$-$\mathcal{T}$ diagram indicating both the EBH and the SBH limits. From bottom to the top, the solid curves indicate the thermodynamic foliations for $r_s=0.4, 0.5, 0.625, 0.75, 0.875, 1$. The green arrow indicates an adiabatic process, going from state (1) to state (2).}}
    \label{fig:STdiagram}
\end{figure}

The determination of these coefficients is based on the contribution of the metric entropy in the Pfaffian. In fact, the B-H entropy relation \cite{Hawking:1974sw,Majumdar:1998}
\begin{equation}\label{eq:B-H}
    S = \frac{k_\mathrm{B} \mathcal{A}_+}{4\ell_\mathrm{p}^2},
\end{equation}
with $k_\mathrm{B}$, $\mathcal{A}_+ = 4\pi r_+^2$ and $\ell_\mathrm{p}$, being respectively the Boltzmann constant, the event horizon area, and the Planck length, implies that $S\equiv S(r_s, r_{\gamma})$. 
Introducing $\tilde{a}\equiv \frac{\pi k_B}{\ell_p^2}$, one can define an entropy function as
\begin{equation}\label{eq:S0}
  \mathcal{S}(r_s, r_{\gamma})  \equiv \frac{S}{\tilde{a}} = r_s^2 \, R_+^2(r_s, r_{\gamma}), 
\end{equation}
where 
\begin{equation}
    \label{eqR} R_+(r_s, r_\gamma)=\frac{r_{\gamma}}{2 r_s} \left(1-\sqrt{1-\frac{4 r_s}{r_{\gamma}}}\right).
\end{equation}
It can be verified that, for any real-valued constant $\lambda$, we have $R_+(\lambda r_s,\lambda r_\gamma) = R_+(r_s,r_\gamma)$. So $R_+$ is homogeneous of degree zero and, therefore, is an "intensive" parameter. 
Since the temperature function $\mathcal{T}$ is an integration factor for the Pfaffian form $\delta Q_{{\rm rev}}=\mathcal{T}\ed \mathcal{S}$, it is obtained from the relation
\begin{equation}\label{eq:T0}
    \mathcal{T} = \left(\frac{\partial \mathcal{S}}{\partial r_s}\right)^{-1}_{r_\gamma} =\frac{1}{r_\gamma} \frac{ \sqrt{1-\frac{4 r_s}{r_{\gamma}}}}{1-\sqrt{1-\frac{4 r_s}{r_{\gamma}}}},
\end{equation}
which is homogeneous of degree $-1$.
{It is informative to demonstrate the mutual behavior of the above thermodynamic parameters in a $\mathcal{S}$-$\mathcal{T}$ diagram (see Fig.~\ref{fig:STdiagram}). As observed from the figure, for a fixed $r_s$, the $\Delta\mathcal{S}>0$ condition corresponds to $\Delta r_\gamma<0$. Therefore, by varying $r_\gamma$ in a particular $r_s$-constant foliation, the system transits towards the Schwarzschild black hole (SBH). It is also straightforward to verify that, going from the state (1) to the state (2) (for which $\mathcal{T}_1<\mathcal{T}_2$), the variable $r_s$ increases. Hence, one can infer that in an adiabatic process, both of the variables $(r_s,r_\gamma)$ increase.}

\begin{figure}[h]
    \centering
    \includegraphics[width=8cm]{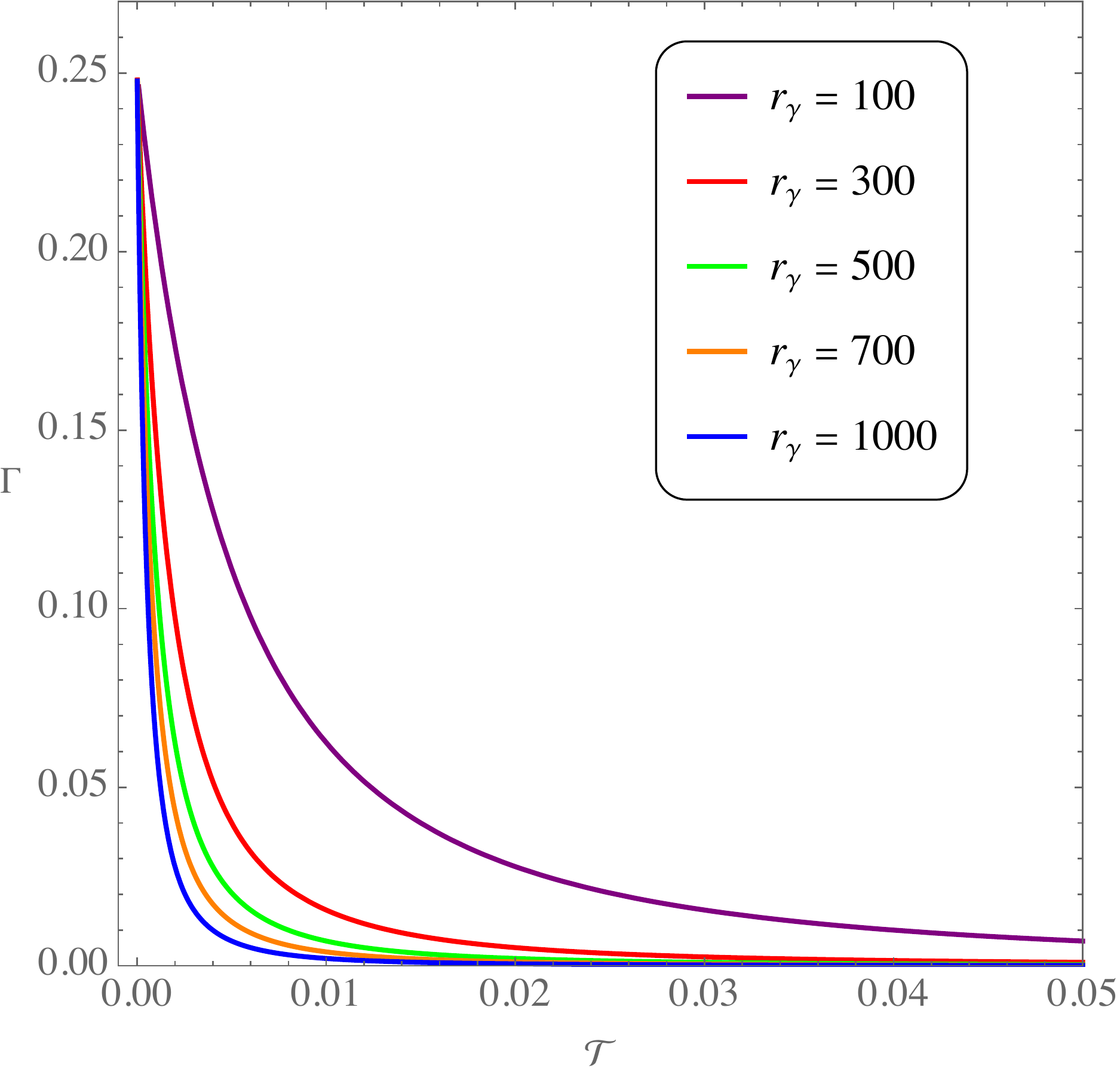}
    \caption{The adiabatic foliations given in terms of the mutual behavior of $\Gamma$ and $\mathcal{T}$, for some values of $r_\gamma$ and varying $r_s$.}
    \label{fig:ST}
\end{figure}

The generalized force is obtained from
\begin{equation}\label{eq:genForce}
    \Gamma=\mathcal{T}\left(\frac{\partial\mathcal{S}}{\partial r_\gamma}\right)_{r_s} = \frac{1}{2}\left(1-\sqrt{1-\frac{4r_s}{r_\gamma}}-\frac{2r_s}{r_\gamma}\right),
\end{equation}
which is an intensive function. Note that, the extremal case corresponds to $\Gamma_e\equiv\Gamma(r_s,r_{\gamma e}) = \frac{1}{4}$, which according to Eq.~\eqref{eq:T0}, is the generalized force for a black hole of zero temperature ($\mathcal{T}=0$).
Thus, performing $(r_s, r_\gamma) \mapsto (\lambda r_s, \lambda r_\gamma)$, we get $\delta Q_{{\rm rev}} \mapsto \lambda \delta Q_{{\rm rev}}$, which means that the Pfaffian form is homogeneous of degree one. In this way, we have an Euler vectorial field, or a Liouville operator, as the infinitesimal generator of the homogeneous transformations
\begin{equation}
    \label{euloper}D\equiv r_s \frac{\partial}{\partial r_s}+r_\gamma \frac{\partial}{\partial r_\gamma}.
\end{equation}
In Fig.~\ref{fig:ST}, the $\Gamma$-$\mathcal{T}$ diagram has been plotted for several fixed values of $r_\gamma$, indicating that the raise in $r_\gamma$ (i.e. the decrease in $\gamma$), increases the negative slope of the curves towards the asymptot, while $\mathcal{T}$ is increasing. 

\section{The adiabatic-isoareal processes and the extremal limit}\label{sec:Extremal}

The correct construction of the thermodynamic manifold leads us to the study of the adiabatic processes that represent the foliations of this manifold. Thus, we must ensure that this construction is consistent with the usual laws of black hole thermodynamics, and in particular, the status of the third law and the connection between extremal and non-extremal states.

\subsection{Thermodynamic limit}\label{subsec:ThermoLimit}

We start by analyzing the surface generated by the extremal states and its implication to the construction of the entire thermodynamic manifold.
In fact, if $\mathbf{r}_s\equiv r_s^e=\Gamma_e r_\gamma$ is the extremal value of $r_s$, then the area of the extremal states can be written as
\begin{equation}
    \label{areext}\mathcal{A}_e=4\pi (r_+^e)^2=4 \pi (r_{s}^e R_+^e)^2=16\pi \mathbf{r}_s^2,
\end{equation}which implies that
\begin{equation}
    {\rm d}\mathcal{A}_e=\pi r_\gamma {\rm d}r_\gamma,
\end{equation} 
and thus, the isoareal condition d$\mathcal{A}_e=0$ is only satisfied by $r_\gamma=$ const. Therefore, although the transformations between extremal states are adiabatic, they are not isoareal. This indicates that the area-entropy law is not valid for extremal states. We can fix this as it has been established previously by letting $\mathcal{S} = 0$ on this manifold \cite{Teitelboim:1994,Carroll:2009}. Despite this, we can also try using a different criterion, as proposed by Lemos  in lower-dimensional gravity \cite{Lemos:2017aol}. For now, we consider that the extremal sub-manifold constitutes the boundary of the thermodynamic manifold, which will be studied later.

As discussed in the previous section, the Pfaffian form is responsible for constructing the physically accepted thermodynamic manifold. In this sense, the Cauchy problem 
\Qr$=0$ generates the non-extremal isentropic (i.e. adiabatic and reversible) sub-manifolds of the thermodynamic foliations \cite{Belgiorno:2002iv}. Let us rewrite the Pfaffian \eqref{eq:1stLaw} as
\begin{equation}\label{eq:1stLawxyz}
    \delta Q_{\mathrm{rev}} = \frac{\ed x}{2\sqrt{x}}
    -\frac{\Gamma(x,y)}{\Gamma_e}\frac{\ed y}{2\sqrt{y}},
\end{equation}
by defining $x\equiv r_s^2$ and $y\equiv\Gamma_e^2 r_\gamma^2$. After some arrangements, these changes of variables yield
\begin{equation}\label{eq:genForcexy}
    \frac{\Gamma(x,y)}{\Gamma_e}=\left(1-\sqrt{1-\sqrt{\frac{x}{y}}}\,\right)^2.
\end{equation}
This way, the Cauchy problem leads to the differential equation
\begin{equation}\label{eq:Cauchy0}
    \frac{\ed y}{\ed x}=\frac{\sqrt{\frac{y}{x}}}{\left(1-\sqrt{1-\sqrt{\frac{x}{y}}}\,\right)^2}=F(x,y),
\end{equation}
that mandates the condition $x<y$ on the thermodynamic manifold.
Note that, $y(x)=x$ is the solution to the above equation which means that extremal states are adiabatically interconnected. 
According to our discussions, this is in conflict with the statement of the second law and its connection with the area (i.e. the B-H) formula. But they can still be reconciled if we consider that both varieties are disconnected, and thus, the third law is also preserved.

The thermodynamic manifold is, therefore, composed by the two mutually inaccessible sub-manifolds, inferred from the following problems for $(x_0,y_0)$ being an initial thermodynamic state:
\begin{itemize}
    \item On the $\mathcal{T}\neq0$ sub-manifold,
    \begin{subequations}\label{eq:Cauchy1}
    \begin{align}
   & \frac{\ed y}{\ed x} = F(x,y),\\
   & y(x_0) = y_0 > x_0.
      \end{align}
\end{subequations}

\item On the $\mathcal{T}=0$ sub-manifold (extremal limit),
    \begin{subequations}\label{eq:Cauchy2}
    \begin{align}
   & \frac{\ed y}{\ed x} = F(x,y),\\
   & y(x_0) = y_0 = x_0.
      \end{align}
\end{subequations}
\end{itemize}
The Cauchy problem \eqref{eq:Cauchy0} has the general solutions (see appendix \ref{app:A})
\begin{equation}
    \label{coffunc}y_i(x; x_0, y_0)=\frac{x}{\left[1-\left(\alpha_i \sqrt{\frac{x}{x_0}} -1\right)^2\right]^2},
\end{equation}where the $\alpha_i$'s depend on the initial condition by

\begin{subequations}\label{eq:y(x)0}
    \begin{align}
     & \alpha_1(x_0, y_0)\equiv \alpha_1 = 1+\sqrt{1+\sqrt{\frac{x_0}{y_0}}}\label{eq:y(x)0a},\\
     & \alpha_2(x_0, y_0)\equiv \alpha_2 = 1+\sqrt{1-\sqrt{\frac{x_0}{y_0}}}\label{eq:y(x)03},\\
     & \alpha_3(x_0, y_0)\equiv \alpha_3 = 1-\sqrt{1-\sqrt{\frac{x_0}{y_0}}}\label{eq:y(x)0c},\\
     & \alpha_4(x_0, y_0)\equiv \alpha_4 = 1-\sqrt{1+\sqrt{\frac{x_0}{y_0}}}\label{eq:y(x)0b}.
    \end{align}
\end{subequations}
Since $y_0>x_0$, the hierarchy order for this parameters is $\alpha_1>\alpha_2>\alpha_3>\alpha_4$. It is interesting to see the particular cases obtained from some initial states. Firstly, if the initial state is a pure quintessence (i.e. $x_0=0$), then $\alpha_1=\alpha_2=2$ and $\alpha_3=\alpha_4=0$, which implies that $y_1(x)=y_2(x)=0$ for all $y_0$, and remains in that state indefinitely. This is while $y_3(x)$ and $y_4(x)$ diverge. Furthermore, if the initial state corresponds to the extremal case, then this functions take the forms
\begin{subequations}\label{eq:y(x)ext}
\begin{align}
 &   y_1^{e}(x) = \frac{x}{\left[1-\left(\sqrt{\frac{x}{x_0}}\left(\sqrt{2}+1\right)-1\right)^2\right]^2},\label{eq:y(x)exta}\\
 &   y_2^{e}(x) = \frac{x}{\left[1-\left(\sqrt{\frac{x}{x_0}}-1\right)^2\right]^2},\label{eq:y(x)extb}\\
 &   y_3^{e}(x) = y_2^{e}(x),\label{eq:y(x)extc}\\
 &   y_4^{e}(x) = \frac{x}{\left[1-\left(\sqrt{\frac{x}{x_0}}\left(\sqrt{2}-1\right)+1\right)^2\right]^2},\label{eq:y(x)extd}
\end{align}
\end{subequations}
which, obviously, must be removed from the set of permissible (physical) solutions.

We can now summarize the properties of these functions considering the following statements:\\

\underline{Proposition 1:}
    {\it Each function is finite at $x=0$ with a value
    \begin{equation}\label{eq:prop_1}
    y_{i0}\equiv \lim_{x\rightarrow 0^+}y_i(x)=\frac{x_0}{4 \alpha_i^2},
    \end{equation}
and thus, $0<y_{10}< y_{20}<y_{30}<y_{40}.$}\\

\underline{Proposition 2:}
    {\it Each function diverges at
    \begin{equation}\label{eq:prop_2}
    x_{\infty}^{(i)}=\frac{4 x_0}{\alpha_i^2},
    \end{equation}
where $0<x_{\infty}^{(1)}<x_{\infty}^{(2)}<x_{\infty}^{(3)}<x_{\infty}^{(4)}$.}\\

\underline{Proposition 3:} {\it For all $x_0\leq y_0$,  the condition 
\begin{equation}\label{eq:prop_3}
x_{\infty}^{(1)}<x_0< x_{\infty}^{(2)}
\end{equation}
holds.}\\

\begin{figure}[h]
    \centering
    \includegraphics[width=8cm]{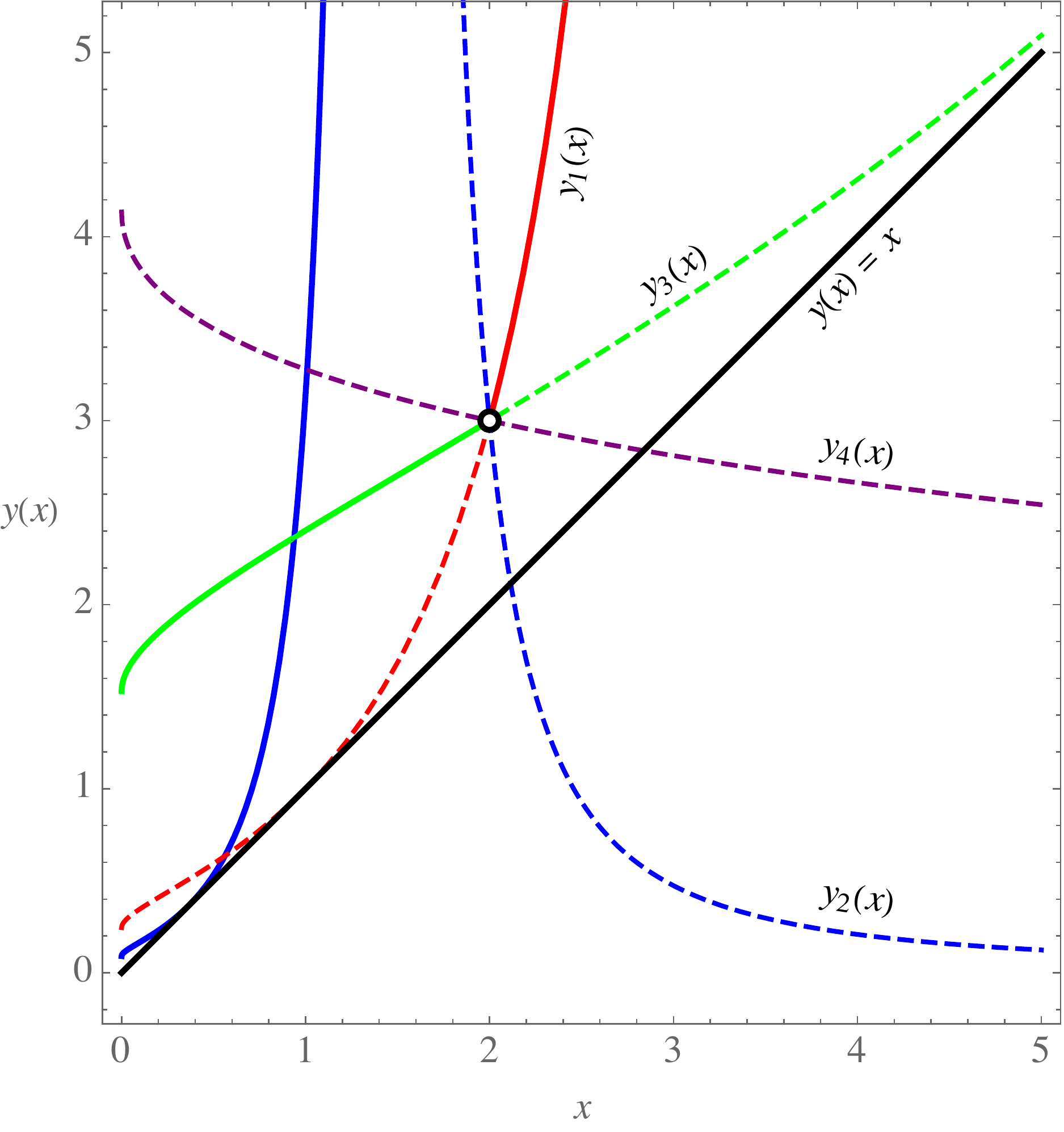}
    ~(a)
     \includegraphics[width=8cm]{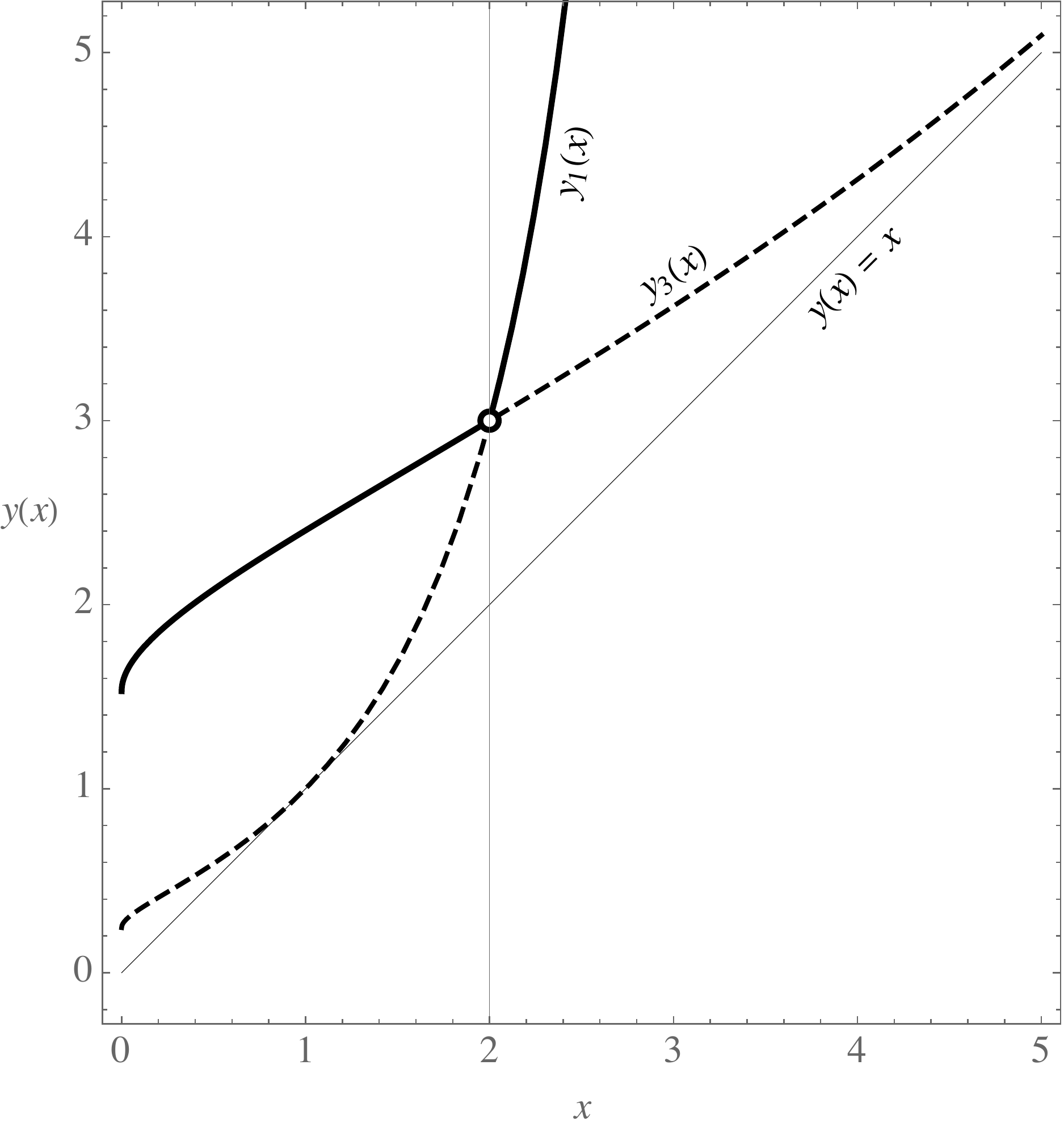}
     ~(b)
    \caption{(a) The branches of $y(x)$ for $(x_0,y_0)=(2, 3)$, shown as a single blank point of intersection, plotted together with the extremal limit (the straight line). The physically accepted paths are shown with solid curves, whereas the dashed curves are not allowed. (b) The physically accepted paths shown in a single diagram.}
    \label{fig:y}
\end{figure}

\begin{figure}[h]
    \centering
    \includegraphics[width=8cm]{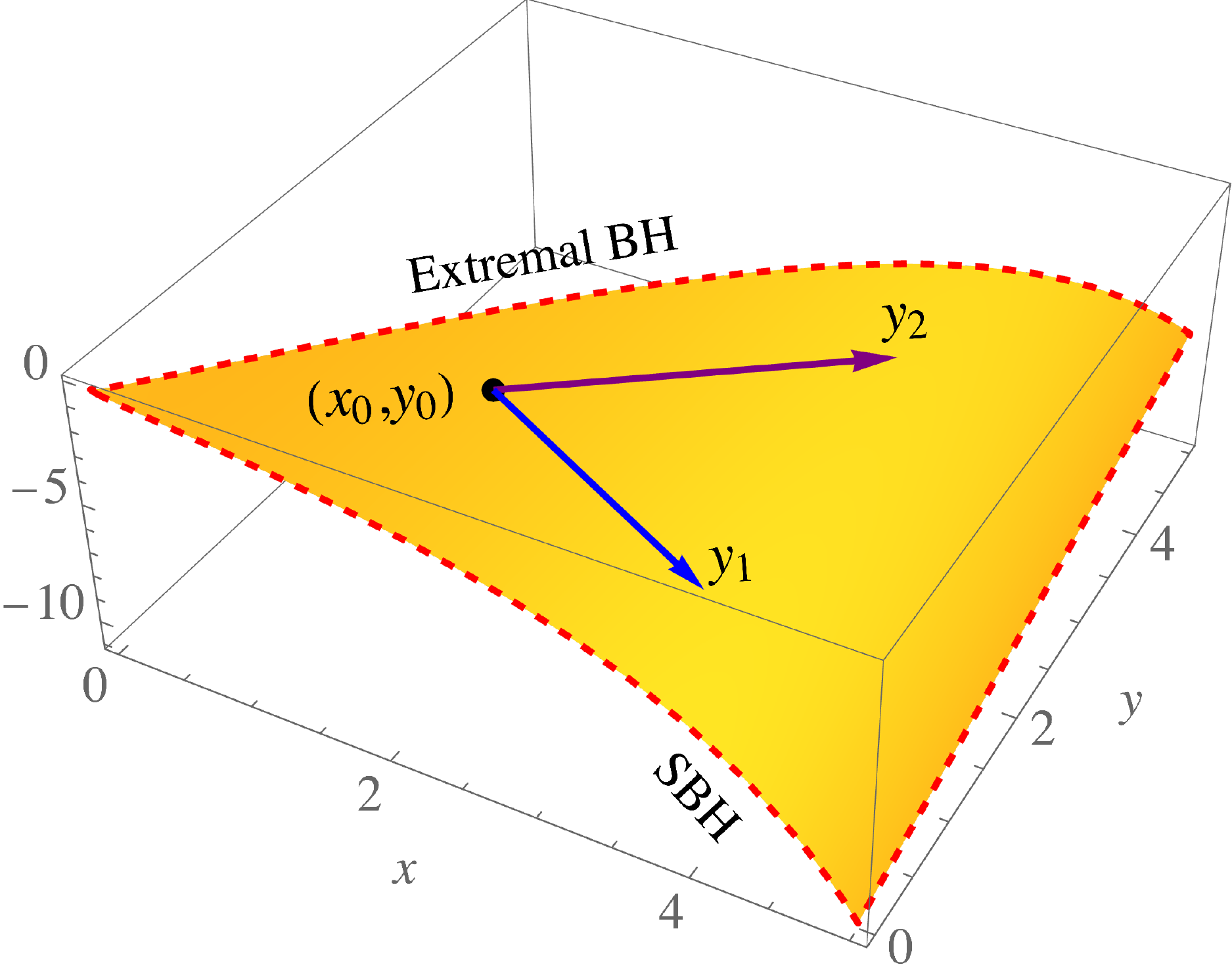}
    \caption{The adiabatic surface constructed by the solution \eqref{eq:y(x)0}. The extremal (i.e. $\mathcal{T}=0$) and the SBH limits have been also indicated.}
    \label{fig:adiabaticSurface}
\end{figure}

\underline{Proposition 4:} Each function intersects the straight line of the extremal states (i.e. $y=x$), at
\begin{subequations}\label{eq:prop_4}
\begin{align}
    & x_{e}^i=\frac{x_0 (\sqrt{2}+1)^2}{\alpha_i^2},\label{eq:prop_4a}\\
    & x_{e'}^i=\frac{x_0}{\alpha_i^2}.\label{eq:prop_4b}
\end{align}
\end{subequations}
In Fig.~\ref{fig:y}, the solutions in Eqs.~\eqref{eq:y(x)0} have been plotted for a particular initial condition, where the extremal limit $y(x)=x$ has been distinctively shown. 
The physically acceptable branches, are however, those that respect the correct changes of the thermodynamics coordinates. Accordingly, the permitted paths are those that allow for simultaneous raise of $y(x)$ and $x$ for the values $x<x_0$. This is while for $x>x_0$, the adiabatic paths should respect a raise in $y(x)$ with a reduction in $x$.
The corresponding adiabatic surface that corresponds to the Cauchy problem has been also plotted in Fig.~\ref{fig:adiabaticSurface}.

\section{The heat capacity and the free energies}\label{sec:HeatCap}

\begin{figure}[h]
    \centering
    \includegraphics[width=8cm]{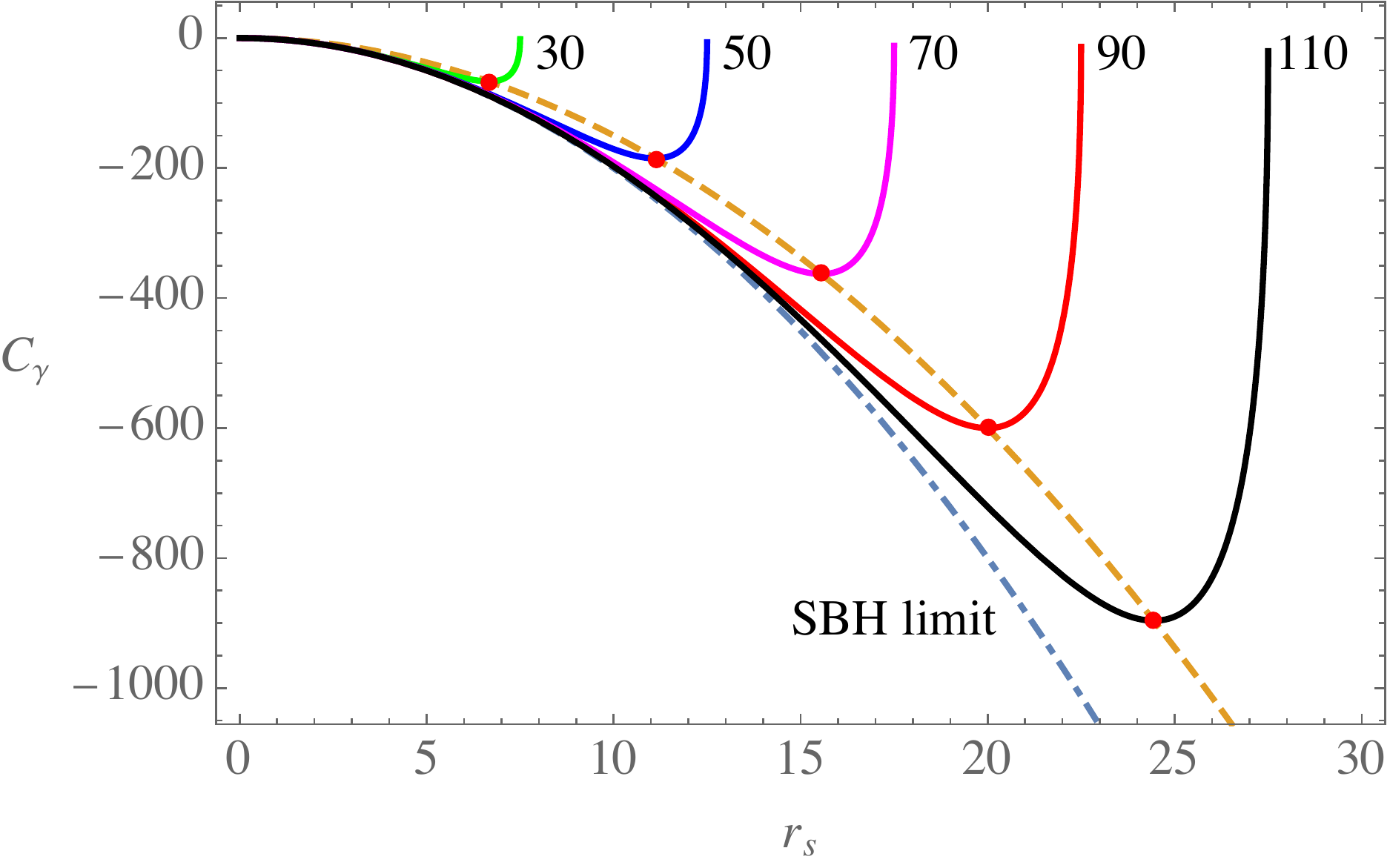}
    \caption{The behavior of $C_\gamma$ plotted for $r_{\gamma} = 30, 50, 70, 90$ and $110$. The dashed curve connects the minimum values $C_\gamma^{\mathrm{m}}$, whereas the dotted-dashed curve corresponds to $C_\gamma^{\mathrm{Sch}}$.} 
    \label{fig:Heat_capacity_1}
\end{figure}

\begin{figure}[h]
    \centering
    \includegraphics[width=8cm]{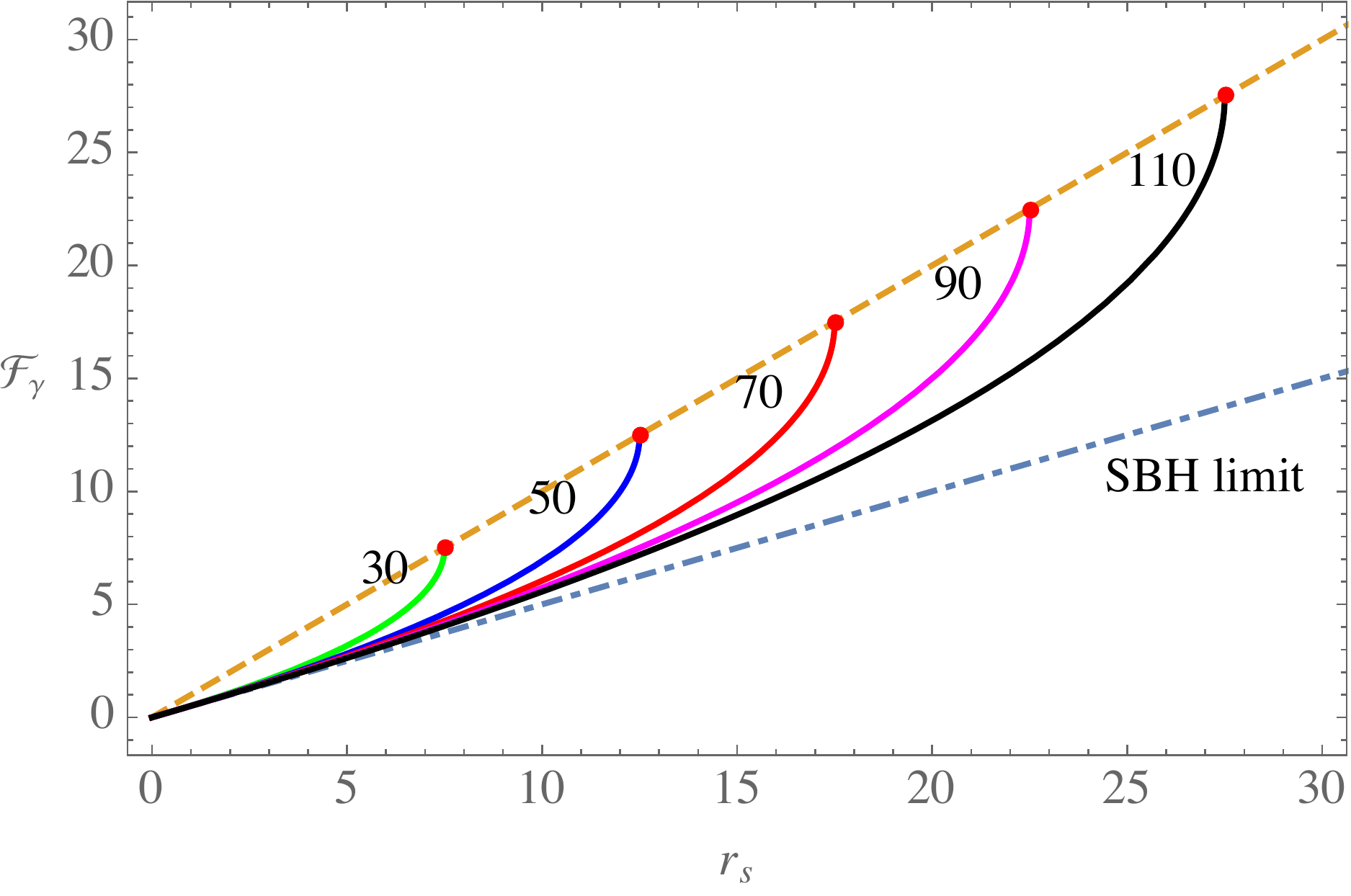}
    \caption{The behavior of $\mathcal{F}_{\gamma}$ plotted for $r_{\gamma} = 30, 50, 70, 90$ and $110$. The dashed line connects the maximum values $\mathcal{F}_{\gamma}^{\mathrm{M}}$, whereas the dotted-dashed line corresponds to $\mathcal{F}_{\gamma}^{\mathrm{Sch}}$. }
    \label{fig:Frgamma}
\end{figure}

\begin{figure}[h]
    \centering
    \includegraphics[width=8cm]{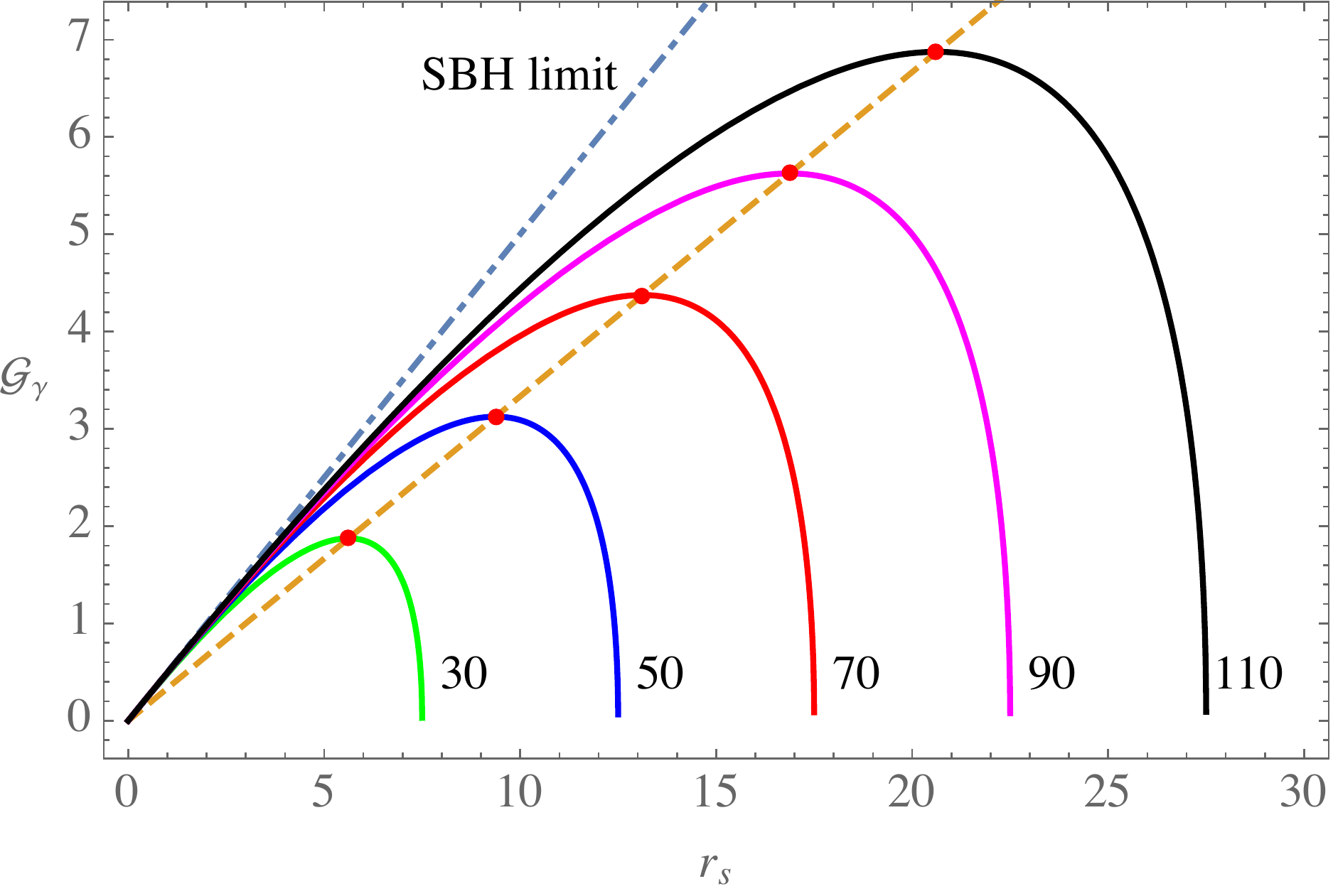}
    \caption{The behavior of $\mathcal{G}_{\gamma}$ plotted for $r_{\gamma} = 30, 50, 70, 90$ and $110$. The dashed line connects the maximum values $\mathcal{G}_{\gamma}^{\mathrm{M}}$, whereas the dotted-dashed line corresponds to $\mathcal{G}_{\gamma}^{\mathrm{Sch}}$.}
    \label{fig:Grgamma}
\end{figure}

The behavior of a thermodynamic system depends strongly on the extent, to which, it can absorb heat. Essentially, we study this heat capacity by keeping $r_\gamma=\mathrm{const.}$, and thus
\begin{equation}\label{eq:Heat_c_1}
    C_\gamma = \mathcal{T} \left( \frac{\partial  \mathcal{S}}{\partial \mathcal{T}}\right)_{r_{\gamma}} =\left( \frac{\partial  r_{s}}{\partial \mathcal{T}}\right)_{r_{\gamma}} = \left( \frac{\partial  \mathcal{T}}{\partial r_{s}}\right)^{-1}_{r_{\gamma}}.
\end{equation}
{Assuming the thermal equilibrium between the black hole and its environment, then by applying Eq.~\eqref{eq:T0} in Eq.~\eqref{eq:Heat_c_1}, we get
\begin{equation}\label{eq:Heat_c_1_explicit}
    C_\gamma = -\frac{r_{\gamma}^{2}}{2} \left(1 - \sqrt{1 - \frac{4 r_{s}}{r_{\gamma}}}\right)^{2} \sqrt{1 - \frac{4 r_{s}}{r_{\gamma}}},
\end{equation}
that reaches its minimum 
\begin{equation}
C_\gamma^{\mathrm{m}}(r_s^\mathrm{m},r_\gamma)\equiv C_\gamma^\mathrm{m} = -\frac{3\left(r_s^\mathrm{m}\right)^2}{2},
    \label{eq:Crgamma-min}
\end{equation}
at $r_s^\mathrm{m} = \frac{2r_\gamma}{9}$. It is of worth mentioning that, the heat capacity of this black hole in the context of the generalized uncertainty principle (GUP) has been calculated in Ref.~\cite{Lutfuoglu:2021}, which reads as
\begin{equation}
    \mathfrak{C}_{\gamma} = -\frac{\beta}{4}\frac{(1-\frac{2r_+}{r_\gamma})\sqrt{1-\frac{\beta}{4r_+^2}}}{(1-\frac{4r_+}{r_\gamma})\left(1-\sqrt{1-\frac{\beta}{4r_+^2}}\right)+\frac{\beta}{2r_+r_\gamma}},
    \label{eq:CGUP}
\end{equation}
where $0<\beta<1$ is the deformation parameter proportional to the Planck length, and accounts for the generalization of the Heisenberg uncertainty principle (HUP)\footnote{Note that, there is also a $\pi$ factor included in Eq.~\eqref{eq:CGUP}, which comes from the authors' version of definition of the HUP.}. Taking into account the expression of $r_+$ in Eq.~\eqref{eq:rH_sB}, it can be seen by direct calculation, that $C_\gamma\rightarrow\mathfrak{C}_{\gamma}$ in the case of $\beta\rightarrow0$.} In Fig.~\ref{fig:Heat_capacity_1}, the behavior of $C_\gamma$ has been plotted for different values of $r_\gamma$. Each of the resultant branches corresponding to fixed values of $r_\gamma$, has a minimum that satisfy the condition $\frac{r_s^{\mathrm{m}}}{r_\gamma} = \frac{2}{9}$.
In the case of $\gamma \rightarrow 0$ (or $r_{\gamma} \rightarrow \infty)$, we have that
\begin{equation}\label{eq:limit_heat_capacity}
    C_\gamma \approx -2 r_{s}^2\left(1 - \frac{2 r_{s}}{r_{\gamma}}\right) \cong -2 r_{s}^2 = C_\gamma^{\mathrm{Sch}},
\end{equation}
which is the heat capacity for the SBH. Note that, the negativity of $C_\gamma$, as seen in Fig.~\ref{fig:Heat_capacity_1}, implies that the black hole gets hotter as it radiates. Furthermore, one can observe that $C_\gamma$ is continuous in the range $0 < r_{s} < \frac{r_{\gamma}}{4}$, which indicates the absence of any kind of phase transition. It is also fruitful to calculate the free energies associated with this heat capacity. According to the notions applied here and in Ref.~\cite{Molina:2021}, the Helmholtz free energy is give by $\mathcal{F}=r_s-\mathcal{T}\mathcal{S}$, that using Eqs.~\eqref{eq:S0} and \eqref{eq:T0},  yields 
\begin{equation}
\mathcal{F}_{\gamma}=\frac{r_\gamma}{4}\left(1-\sqrt{1-\frac{4r_s}{r_\gamma}}\right),
\label{eq:Helmholtz}
\end{equation}
which is presented for the case of $r_\gamma=$ const. In Fig.~\ref{fig:Frgamma}, the behavior of the $\mathcal{F}_{\gamma}$ has been ramified for the same values of $r_\gamma$ as those used in Fig.~\ref{fig:Heat_capacity_1}.
As it is inferred from the figure, the Helmholtz free energy exhibits the upper limit $\mathcal{F}_{\gamma}^{\mathrm{M}}=r_s$, for each of the branches. Furthermore, it respects the limit $\mathcal{F}_{\gamma} \rightarrow \frac{r_s}{2} = \mathcal{F}_{\gamma}^\mathrm{Sch}$ for very large $r_\gamma$, indicating its value for the SBH. This way, and as inferred from Fig.~\ref{fig:Frgamma}, the Helmholtz free energy is confined within the domain $\mathcal{F}_{\gamma}^\mathrm{Sch}\leq\mathcal{F}_{\gamma}\leq\mathcal{F}_{\gamma}^{\mathrm{M}}$.
Moreover, the Gibbs free energy can be calculated by means of the relation $\mathcal{G}=r_s-\mathcal{T}\mathcal{S}-\Gamma r_\gamma$  \cite{Molina:2021}, which by means of Eqs.~\eqref{eq:S0}, \eqref{eq:T0} and \eqref{eq:genForce}, provides
\begin{equation}
\mathcal{G}_{\gamma} = r_s-\frac{r_\gamma}{4}\left(1-\sqrt{1-\frac{4r_s}{r_\gamma}}\right),
    \label{eq:Gibbs}
\end{equation}
for constant values of $r_\gamma$. This quantity has a maximum of
\begin{equation}
\mathcal{G}_{\gamma}^{\mathrm{M}}(r_s^\mathrm{M},r_\gamma)\equiv \mathcal{G}_{\gamma}^\mathrm{M} = \frac{r_s^\mathrm{M}}{3},
    \label{eq:Crgamma-min}
\end{equation}
for $r_s^\mathrm{M} = \frac{3r_\gamma}{16}$. The behavior of $\mathcal{G}_{\gamma}$ has been shown in Fig.~\ref{fig:Grgamma}.
Either of the branches in the diagram, possesses a maximum satisfying the relation $\frac{r_s^M}{r_\gamma}=\frac{3}{16}$. Note that, as expected, for very large values of $r_\gamma$, we have $\mathcal{G}_{\gamma}\rightarrow\frac{r_s}{2}=\mathcal{G}_{\gamma}^\mathrm{Sch}$, that corresponds to its value for the SBH. So, in the absence of quintessence, both of the Helmholtz and Gibbs free energies will lead to the same value, which is that of the SBH.

\section{Conclusions}\label{sec:conclusions}

The thermodynamic features of black holes, beside being of interest on their own as the physical tools for attributing sensible physical phenomena to extremally gravitating systems, can also be important regarding modern cosmology, when these systems are associated with some dark fields. Basically, after redefining the causality of the spacetime by introducing new horizons, such cosmological components also affect the thermodynamics of the black hole through the extra terms they induce to the spacetime metric. In this paper, we showed that the quintessential dark field surrounding a SBH, can be regarded as a thermodynamic coordinate, and be used as a parameter together with the black hole mass. Applying the Pfaffian form of the infinitesimal heat
exchange reversibly, we calculated the geometric entropy $\mathcal{S}$ and temperature $\mathcal{T}$ in terms of these coordinates. Accordingly, we foliated the  $\mathcal{S}$-$\mathcal{T}$ curves for different black hole masses, and showed that during the adiabatic processes, both of the thermodynamic coordinates increase. This inference is crucial for this particular black hole, because we saw further that the solutions to the Cauchy problem which are given in the context of isoareal (adiabatic) processes, lead to several different paths on the thermodynamic manifold, that not all of them are physically meaningful. Accordingly, one needs to choose, among these paths, those that rely on the increase of both of the parameters. We also showed the acceptable paths in two-dimensional and three-dimensional plots. Furthermore, we used the formerly derived thermodynamic quantities, to calculate the free energies associated with the black hole horizon. We found that for constant values of the cosmological component, the free energies are limited, and in the absence of the cosmological term, both of them lead correctly to the same value for the SBH. The discussion presented in this paper, can be generalized to the rotating counterpart of the black hole which may be accounted for some astrophysical support. This investigation is left for future studies.

\section*{Acknowledgements}
M. Fathi acknowledges Universidad de Santiago de Chile for financial support through the Proyecto POSTDOCDICYT, C\'{o}digo 042331 CM-Postdoc. J.R. Villanueva was partially supported by the Centro de Astrof\'isica de Valpara\'iso (CAV).

\appendix

\section{Derivation of the solution to the Cauchy problem}\label{app:A}

Applying the change of variable $z\doteq \sqrt{\frac{x}{y}}$, the differential equation \eqref{eq:Cauchy0} can be rewritten as
\begin{equation}\label{eq:A1}
    \frac{1}{z^2}-\frac{2x}{z^3}\frac{\ed z}{\ed x}=\frac{1}{z(1-\sqrt{1-z})^2},
\end{equation}
which can be recast as
\begin{equation}\label{eq:A2}
    \frac{\ed w}{\ed x}=\frac{1+w}{2x},
\end{equation}
using the second change of variable $w\doteq\sqrt{1-z}$. Considering the initial states $(x_0,w_0)$, the above equation results in the solution
\begin{equation}\label{eq:A3}
    \frac{\sqrt{x}}{1+w}=\bar{c}_1,
\end{equation}
where $\bar{c}_1=\frac{\sqrt{x_0}}{1+w_0}$.
After doing some algebraic arrangements, and taking into account the applied changes of available, we finally get to the solutions in Eqs.~\eqref{eq:y(x)0}.

\bibliographystyle{ieeetr}
\bibliography{biblio_v1.bib}

\begin{thebibliography}{10}

\bibitem{Bardeen:1973gs}
J.~M. Bardeen, B.~Carter, and S.~Hawking, ``{The Four laws of black hole
  mechanics},'' {\em Commun. Math. Phys.}, vol.~31, pp.~161--170, 1973.

\bibitem{Bekenstein:1972}
J.~D. Bekenstein, ``Black holes and the second law,'' {\em {Lett. Nuovo
  Cimento}}, vol.~4, pp.~737--740, Aug 1972.

\bibitem{Bekenstein:1973}
J.~D. Bekenstein, ``Black holes and entropy,'' {\em Phys. Rev. D}, vol.~7,
  pp.~2333--2346, Apr 1973.

\bibitem{Bekenstein:1974}
J.~D. Bekenstein, ``Generalized second law of thermodynamics in black hole
  physics,'' {\em Phys. Rev. D}, vol.~9, pp.~3292--3300, Jun 1974.

\bibitem{Bekenstein:1975}
J.~D. Bekenstein, ``Statistical black hole thermodynamics,'' {\em Phys. Rev.
  D}, vol.~12, pp.~3077--3085, Nov 1975.

\bibitem{Hawking:1974sw}
S.~Hawking, ``{Particle creation by black holes},'' {\em Commun. Math. Phys.},
  vol.~43, pp.~199--220, 1975.
\newblock [Erratum: Commun.Math.Phys. 46, 206 (1976)].

\bibitem{Teitelboim:1994}
C.~Teitelboim, ``{Action and entropy of extreme and nonextreme black holes},''
  {\em Phys. Rev. D}, vol.~51, p.~4315, 1995.
\newblock [Erratum: Phys.Rev.D 52, 6201 (1995)].

\bibitem{Carroll:2009}
S.~M. Carroll, M.~C. Johnson, and L.~Randall, ``Extremal limits and black hole
  entropy,'' {\em J. High Energy Phys.}, vol.~2009, pp.~109--109, Nov. 2009.

\bibitem{Unruh:1981}
W.~G. {Unruh}, ``{Experimental Black-Hole Evaporation?},'' {\em Physical Review
  Letters}, vol.~46, pp.~1351--1353, May 1981.

\bibitem{Novello:2002}
M.~Novello, M.~Visser, and G.~Volovik, {\em Artificial Black Holes}.
\newblock World Scientific, 2002.

\bibitem{Schutzhold:2005}
R.~{Sch{\"u}tzhold} and W.~G. {Unruh}, ``{Hawking Radiation in an
  Electromagnetic Waveguide?},'' {\em Physical Review Letters}, vol.~95,
  p.~031301, July 2005.

\bibitem{Carusotto:2008}
I.~Carusotto, S.~Fagnocchi, A.~Recati, R.~Balbinot, and A.~Fabbri, ``Numerical
  observation of {Hawking} radiation from acoustic black holes in atomic
  {Bose}-{Einstein} condensates,'' {\em New Journal of Physics}, vol.~10,
  p.~103001, Oct. 2008.

\bibitem{Belgiorno:2010}
F.~Belgiorno, S.~L. Cacciatori, M.~Clerici, V.~Gorini, G.~Ortenzi, L.~Rizzi,
  E.~Rubino, V.~G. Sala, and D.~Faccio, ``Hawking radiation from ultrashort
  laser pulse filaments,'' {\em Phys. Rev. Lett.}, vol.~105, p.~203901, Nov
  2010.

\bibitem{Weinfurtner:2011}
S.~{Weinfurtner}, E.~W. {Tedford}, M.~C.~J. {Penrice}, W.~G. {Unruh}, and G.~A.
  {Lawrence}, ``{Measurement of Stimulated Hawking Emission in an Analogue
  System},'' {\em Phys. Rev. Lett.}, vol.~106, p.~021302, Jan. 2011.

\bibitem{Castelvecchi:2016}
D.~Castelvecchi, ``Artificial black hole creates its own version of {Hawking}
  radiation,'' {\em Nature}, vol.~536, pp.~258--259, Aug. 2016.

\bibitem{Steinhauer:2016}
J.~Steinhauer, ``Observation of quantum {Hawking} radiation and its
  entanglement in an analogue black hole,'' {\em Nature Physics}, vol.~12,
  pp.~959--965, Oct. 2016.

\bibitem{Lima:2019}
C.~A.~U. Lima, F.~Brito, J.~A. Hoyos, and D.~A.~T. Vanzella, ``Probing the
  {Unruh} effect with an accelerated extended system,'' {\em Nature
  Communications}, vol.~10, p.~3030, Dec. 2019.

\bibitem{Kolobov:2021}
V.~I. Kolobov, K.~Golubkov, J.~R. Muñoz~de Nova, and J.~Steinhauer,
  ``Observation of stationary spontaneous {Hawking} radiation and the time
  evolution of an analogue black hole,'' {\em Nature Physics}, vol.~17,
  pp.~362--367, Mar. 2021.

\bibitem{Hutter:1977}
K.~Hutter, ``The foundations of thermodynamics, its basic postulates and
  implications. {A} review of modern thermodynamics,'' {\em Acta Mechanica},
  vol.~27, pp.~1--54, Mar. 1977.

\bibitem{Neumaier:2007}
A.~{Neumaier}, ``{On the foundations of thermodynamics},'' {\em arXiv
  e-prints}, p.~arXiv:0705.3790, May 2007.

\bibitem{caratheodory09}
C.~Carath\'eodory, ``{Untersuchungen \"uber die Grundlagen der
  Thermodynamik.},'' {\em Math. Ann.}, vol.~67, pp.~355--386, 1909.

\bibitem{Landsberg:1956}
P.~T. Landsberg, ``Foundations of thermodynamics,'' {\em Rev. Mod. Phys.},
  vol.~28, pp.~363--392, Oct 1956.

\bibitem{Antoniou:2002}
I.~E. Antoniou, ``[{No} title found],'' {\em Foundations of Physics}, vol.~32,
  no.~4, pp.~627--641, 2002.

\bibitem{Gibbs:1949}
J.~Gibbs, {\em Thermodynamics, Vol. 1}.
\newblock Yale University Press, New Haven, CT, 1948.

\bibitem{Born:1949}
M.~Born, {\em Natural Philosophy of Cause and Chance}.
\newblock Clarendon Press, Oxford, 1949.

\bibitem{Quevedo:2007}
H.~Quevedo, ``Geometrothermodynamics,'' {\em Journal of Mathematical Physics},
  vol.~48, no.~1, p.~013506, 2007.

\bibitem{Belgiorno:2002}
F.~{Belgiorno}, ``{Homogeneity as a bridge between Carath{\'e}odory and
  Gibbs},'' {\em arXiv e-prints}, pp.~math--ph/0210011, Oct. 2002.

\bibitem{Belgiorno:2002iv}
F.~Belgiorno, ``{Black hole thermodynamics in Caratheodory's approach},'' {\em
  Phys. Lett. A}, vol.~312, pp.~324--330, 2003.

\bibitem{Belgiorno:2002iw}
F.~Belgiorno, ``{Quasihomogeneous thermodynamics and black holes},'' {\em J.
  Math. Phys.}, vol.~44, pp.~1089--1128, 2003.

\bibitem{Belgiorno:2003a}
F.~{Belgiorno}, ``{Notes on the third law of thermodynamics: I},'' {\em J.
  Phys. A: Math. Gen.}, vol.~36, pp.~8165--8193, Aug. 2003.

\bibitem{Belgiorno:2003b}
F.~Belgiorno, ``Notes on the third law of thermodynamics: {II},'' {\em J. Phys.
  A: Math. Gen.}, vol.~36, pp.~8195--8221, jul 2003.

\bibitem{Belgiorno:2004}
F.~Belgiorno and M.~Martellini, ``{Black holes and the third law of
  thermodynamics},'' {\em Int. J. Mod. Phys.}, vol.~D13, pp.~739--770, 2004.

\bibitem{Molina:2021}
{M. Molina and J.R. Villanueva}, ``{On the thermodynamics of the Hayward black
  hole},'' {\em {Class. Quant. Grav.}}, vol.~38, no.~10, p.~105002, 2021.

\bibitem{Fathi:2021EPJC}
M.~Fathi, S.~Lepe, and J.~R. Villanueva, ``Adiabatic analysis of the rotating
  {BTZ} black hole,'' {\em The European Physical Journal C}, vol.~81, p.~499,
  June 2021.

\bibitem{Fathi:2021PLB}
M.~Fathi, M.~Molina, and J.~Villanueva, ``Adiabatic evolution of hayward black
  hole,'' {\em Physics Letters B}, vol.~820, p.~136548, 2021.

\bibitem{JimenezMadrid:2005}
J.~A. Jimenez~Madrid and P.~F. Gonzalez-Diaz, ``{Evolution of a kerr-newman
  black hole in a dark energy universe},'' {\em Grav. Cosmol.}, vol.~14,
  pp.~213--225, 2008.

\bibitem{Jamil:2009}
M.~Jamil, ``{Evolution of a Schwarzschild black hole in phantom-like Chaplygin
  gas cosmologies},'' {\em Eur. Phys. J. C}, vol.~62, pp.~609--614, 2009.

\bibitem{Li:2019}
X.-Q. Li, B.~Chen, and L.-l. Xing, ``{Charged Lovelock black holes in the
  presence of dark fluid with a nonlinear equation of state},'' {\em Eur. Phys.
  J. Plus}, vol.~135, no.~2, p.~175, 2020.

\bibitem{Roy:2020}
R.~Roy and U.~A. Yajnik, ``Evolution of black hole shadow in the presence of
  ultralight bosons,'' {\em Physics Letters B}, vol.~803, p.~135284, 2020.

\bibitem{Xu:2018}
Z.~Xu, X.~Hou, X.~Gong, and J.~Wang, ``Black hole space-time in dark matter
  halo,'' {\em Journal of Cosmology and Astroparticle Physics}, vol.~2018,
  pp.~038--038, sep 2018.

\bibitem{Das:2021}
A.~Das, A.~Saha, and S.~Gangopadhyay, ``Investigation of circular geodesics in
  a rotating charged black hole in the presence of perfect fluid dark matter,''
  {\em Classical and Quantum Gravity}, vol.~38, p.~065015, feb 2021.

\bibitem{Kiselev:2003}
V.~V. Kiselev, ``Quintessence and black holes,'' {\em Classical and Quantum
  Gravity}, vol.~20, pp.~1187--1197, Mar. 2003.

\bibitem{Saadati:2019}
R.~Saadati and F.~Shojai, ``Bending of light in a universe filled with
  quintessential dark energy,'' {\em Phys. Rev. D}, vol.~100, p.~104041, Nov
  2019.

\bibitem{AliKhan:2020}
I.~Ali~Khan, A.~Sultan~Khan, and S.~Islam, ``Dynamics of the particle around de
  sitter–schwarzschild black hole surrounded by quintessence,'' {\em
  International Journal of Modern Physics A}, vol.~35, no.~23, p.~2050130,
  2020.

\bibitem{CFOV21}
V.~H. C\'ardenas, M.~Fathi, M.~Olivares, and J.~R. Villanueva, ``{Probing the
  parameters of a Schwarzschild black hole surrounded by quintessence and cloud
  of strings through four standard astrophysical tests},'' {\em Eur. Phys. J.
  C}, vol.~81, no.~10, p.~866, 2021.

\bibitem{Tharanath:2013}
R.~Tharanath and V.~C. Kuriakose, ``{Thermodynamics} {and} {spectroscopy} {of}
  {Schwarzschild} {black} {hole} {surrounded} {by} {quintessence},'' {\em
  Modern Physics Letters A}, vol.~28, p.~1350003, Feb. 2013.

\bibitem{Ghaderi2016}
K.~Ghaderi and B.~Malakolkalami, ``Thermodynamics of the schwarzschild and the
  reissner–nordström black holes with quintessence,'' {\em Nuclear Physics
  B}, vol.~903, pp.~10--18, 2016.

\bibitem{Ma:2017}
M.-S. Ma, R.~Zhao, and Y.-Q. Ma, ``Thermodynamic stability of black holes
  surrounded by quintessence,'' {\em General Relativity and Gravitation},
  vol.~49, p.~79, June 2017.

\bibitem{Ghosh:2018}
S.~G. Ghosh, S.~D. Maharaj, D.~Baboolal, and T.-H. Lee, ``Lovelock black holes
  surrounded by quintessence,'' {\em The European Physical Journal C}, vol.~78,
  p.~90, Feb. 2018.

\bibitem{Rodrigue:2018}
K.~K.~J. Rodrigue, M.~Saleh, B.~B. Thomas, and T.~C. Kofane, ``Thermodynamics
  phase transition and {Hawking} radiation of the {Schwarzschild} black hole
  with quintessence-like matter and a deficit solid angle,'' {\em General
  Relativity and Gravitation}, vol.~50, p.~52, May 2018.

\bibitem{Nam:2020}
C.~H. Nam, ``Higher dimensional charged black hole surrounded by quintessence
  in massive gravity,'' {\em General Relativity and Gravitation}, vol.~52,
  p.~1, Jan. 2020.

\bibitem{Lutfuoglu:2021}
B.~C. Lütfüoğlu, B.~Hamil, and L.~Dahbi, ``Thermodynamics of {Schwarzschild}
  black hole surrounded by quintessence with generalized uncertainty
  principle,'' {\em The European Physical Journal Plus}, vol.~136, p.~976,
  Sept. 2021.

\bibitem{chandrasekhar39}
S.~Chandrasekhar, {\em An Introduction to the Study of Stellar Structure}.
\newblock Astrophysical monographs, University of Chicago Press, 1939.

\bibitem{Buchdahl:1966}
H.~A. Buchdahl, {\em The Concept of Classical Thermodynamics}.
\newblock Cambridge University Press, Cambridge, 1966.

\bibitem{adkins_1983}
C.~J. Adkins, {\em Equilibrium Thermodynamics}.
\newblock Cambridge University Press, 1983.

\bibitem{Majumdar:1998}
P.~Majumdar, ``{Black hole entropy and quantum gravity},'' {\em Indian J. Phys.
  B}, vol.~73, p.~147, 1999.

\bibitem{Lemos:2017aol}
J.~P.~S. Lemos, M.~Minamitsuji, and O.~B. Zaslavskii, ``{Unified approach to
  the entropy of an extremal rotating BTZ black hole: Thin shells and horizon
  limits},'' {\em Phys. Rev. D}, vol.~96, no.~8, p.~084068, 2017.

\end{thebibliography}

\end{document}